\documentclass[aps, showpacs, pra, superscriptadaddress, eqsecnum, twocolumn
] {revtex4-1}
\usepackage{hyperref}
\usepackage{amsmath}
\usepackage{amsfonts} 
\usepackage{amssymb}
\usepackage{epsfig}
\usepackage{natbib}
\usepackage{epstopdf}
\usepackage{graphicx}
\usepackage[normalem]{ulem}

\usepackage{floatrow}
\usepackage[caption=false,font=Large]{subfig}

\usepackage{amssymb}
\usepackage{amsthm}
\usepackage{color}

\newcommand*{\be}{\begin{equation}}
\newcommand*{\ee}{\end{equation}}
\newcommand*{\bea}{\begin{eqnarray}}
\newcommand*{\eea}{\end{eqnarray}}

\begin{document}

\title{Gyrating solitons in  a necklace of optical waveguides}

 \author{I. V. Barashenkov}
\affiliation{Centre for Theoretical and Mathematical Physics, 
University of Cape Town, Rondebosch 7701, South Africa   and  Joint Institute for Nuclear Research, Dubna, Russia}

 \author{Daniel Feinstein}
 \affiliation{ Keble College, University of Oxford, Parks Road Oxford OX1 3PG, UK  
}

\vspace*{10mm}

\begin{abstract}
We consider  light pulses in a circular array of $2N$ coupled nonlinear optical waveguides. 
The waveguides are either hermitian or alternate gain and loss
in a $\mathcal{PT}$-symmetric fashion.
Simple patterns in the array include a ring of $2N$  pulses travelling abreast,
and a breather --- a string of pulses  where all even and all odd waveguides flash in turn. In addition, 
the structure displays  solitons gyrating around the necklace by switching from one waveguide to the next.
Some  of the  gyrating solitons are stable while other ones are weakly unstable and evolve into gyrating multiflash strings.
By tuning the gain-loss coefficient, the gyration of solitons in a nonhermitian array may be reversed without changing the direction of their translational motion.

\end{abstract}


\maketitle

\section{Introduction}

The uses of nonlinear fibre arrays for the  all-optical signal
processing have been recognised since the late 1980s.
The multiple channel waveguide couplers and multicore fibres
can be utilised for  light switching   \cite{switch,Krolikowski},
power dividing \cite{Hudg}, beam shaping \cite{beam_shaping},
discrete diffraction management \cite{Longhi},
spatial-division multiplexing \cite{multiplex},
coherent beam combination and optical pulse compression
\cite{Aceves}. In recent years interest has been
shifting to  low-dimensional arrays, typically arranged in a ring \cite{ring,Chekh}.
New applications of circular arrays include  vortex switching schemes  \cite{Alexeyev,vortices}
and generation of light beams carrying orbital angular momentum  \cite{OAM}.

Studies of coupled waveguides  have received a new impetus with the advent of the parity-time symmetry.
Originally proposed in the context of the nonhermitian quantum mechanics
 \cite{PT}, the $\mathcal{PT}$-symmetry  proved to furnish a set of rules for the inclusion
 of gain and loss in 
 fiber arrays \cite{PTwaveguide}  and photonic lattices \cite{PTphot}.
The nonhermitian optics  provides light-control opportunities unattainable with traditional set-ups, 
 including
 low-threshold switching \cite{PTwaveguide,SXK}
and unidirectional
invisibility \cite{PTwaveguide,Lin}.

A circular array of waveguides is an ideal platform for the $\mathcal{PT}$-symmetric modification.
An example of such a development is a ring-shaped necklace of $2N$ waveguides
   with alternating gain and loss.
Ref \cite{Liam} has demonstrated that the zero-amplitude state in the $\mathcal{PT}$-symmetric necklace 
    remains stable as long as  $N$ is odd and the gain-loss coefficient  does not exceed a finite threshold.
    The author of Ref \cite{Longhi1} has pointed out then that  the stability in the necklace 
    can be controlled by twisting it about the central axis.
    Further studies concerned stationary modes in 
     a cyclic array of $\mathcal{PT}$ symmetric dimers \cite{Mex},  a  hermitian waveguide ring with 
     a $\mathcal{PT}$-symmetric impurity
     \cite{LKD}, 
and    a  multicore fiber   
    with gain  in the central core and loss in the surrounding ring of waveguides    \cite{Molina}.

\begin{figure}
\begin{center}
\includegraphics[width=\linewidth]{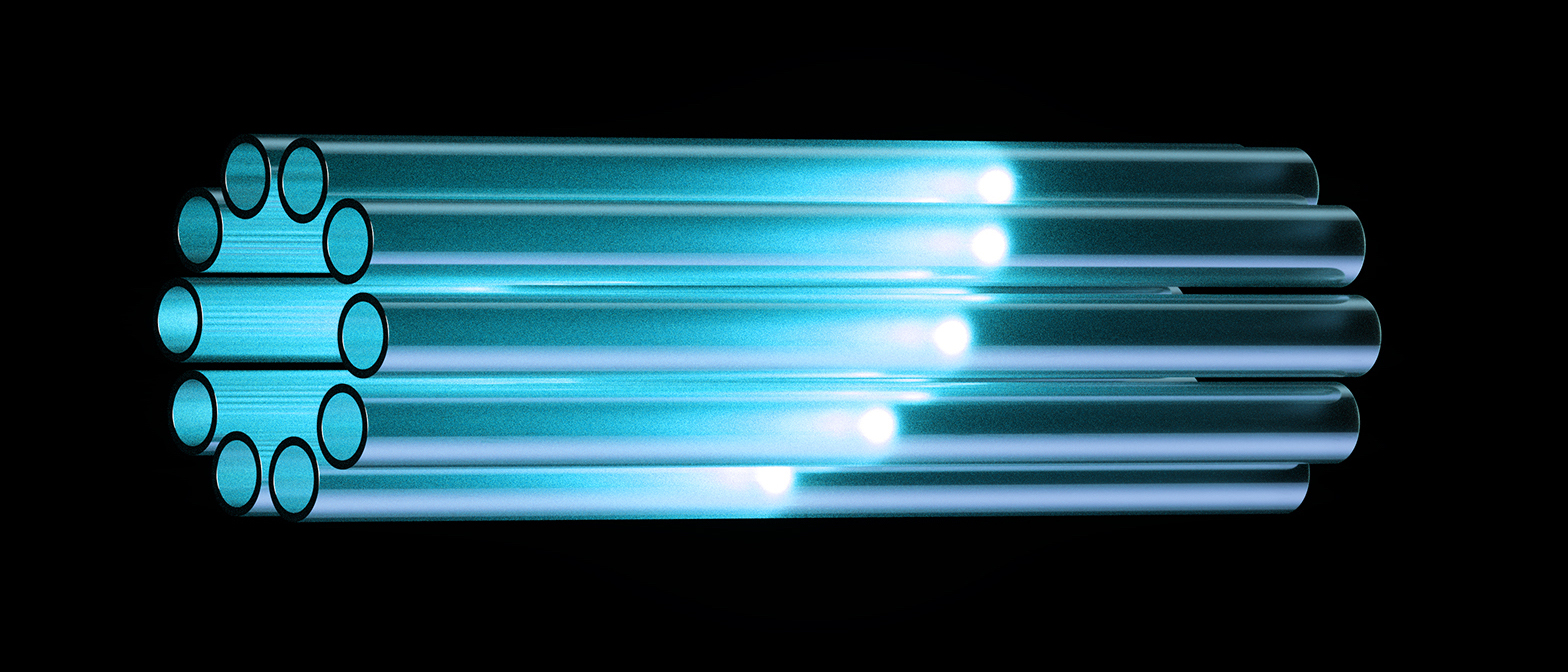}
\end{center}
\caption{A gyrating soliton in the necklace of $2N=10$ waveguides (schematic representation).
 \label{light}}
 \end{figure}

With a few notable exceptions \cite{Chekh}, 
 studies of hermitian and 
 $\mathcal{PT}$-symmetric necklaces have been focussing on  the stationary states of light.
The aim of the present work is to consider short optical  pulses. We
show that the necklace of nonlinear dispersive waveguides ---
with or without gain and loss --- supports solitons of new type. 
As these light pulses propagate along the axis of the multicore fiber, 
they gyrate around the necklace switching from one waveguide to another (Fig \ref{light}).

There are several types of gyrating solitons coexisting in the array of the same number of guides. 
Some of these objects consist of a single pulse that spirals around the necklace;
other ones comprise series of pulses of varied brightness.
 There are solitons with different propagation constants within each of the two varieties.
While the systematic classification of  stability properties of the gyrating solitons is beyond the scope of the
present study,  our analysis indicates that some of these are stable.

The paper is organised as follows. 
In the next section  (section \ref{supermodes}) we classify  linear supermodes in the  nondispersive necklace. These will serve as starting points for the bifurcating nonlinear patterns
(section \ref{rules}). 
In section \ref{sims} we consider constellations of  pulses appearing simultaneously in all $2N$ waveguides
and in sections \ref{breathers} and \ref{Nonuni} discuss solitons oscillating between even and odd subsets of the array.
Solitons whose motion along the fiber is accompanied by their gyration around the necklace, 
 are introduced  in section \ref{Gyro1}. In the subsequent section (\ref{Gyro2}) 
we consider more complex, multiflash, gyrating patterns. Stability and interaction of gyrating solitons are touched upon in section \ref{Dynamics}.
Section \ref{Conclusions} summarises results of this study.

\section{Linear nondispersive waveguides}
\label{supermodes}

The necklace  of waveguides is described by the following system of $2N$ equations written in the reference frame traveling at the common group velocity 
\cite{Mumtaz,Chekh}:
\be
i \partial_z \psi_n +  \partial_\tau^2 \psi_n
+ \psi_{n-1}+\psi_{n+1} +
2|\psi_n|^2\psi_n=2i \Gamma_n \psi_n.
\label{A1}
\ee
 Here $\psi_n$ is the amplitude of the complex mode in the $n$-th core  ($n=1,..., 2N$); 
$z$ measures length along  the device and $\tau$ is a retarded time. 
We are considering waveguides with an anomalous group velocity dispersion and all coefficients
have been normalised to unity.

In the system \eqref{A1} we have assumed that waveguides with gain and loss alternate:
\[
\Gamma_n= (-1)^{n+1} \gamma.
\]
Here $\gamma>0$ is a common gain-loss coefficient. Skipping ahead a bit, 
many of our results will remain valid for the hermitian array, $\gamma=0$. 

The equation \eqref{A1} with $n=1$
 contains an unknown $\psi_0$ 
and the equation with $n=2N$ includes  $\psi_{2N+1}$.
These two variables are defined by virtue of  
 the periodicity condition:
\[
\psi_{n+2N}=\psi_n.
\]

 We start  by examining  the linear nondispersive limit of \eqref{A1} which results from dropping the nonlinearity
 and  time derivative $\partial_\tau^2 \psi_n$.
 Assuming a 
  separable solution of the form 
 $\psi_n= v_n e^{i \beta z}$,
 the coefficients $v_n$ comprise an eigenvector ${\vec v}=(v_1, v_2, ..., v_{2N})^T$
 of the matrix $\mathcal L$:
 \[
 \mathcal L {\vec v} = \beta {\vec v},
 \]
 where 
 \be
 \mathcal L_{nm}= \delta_{n, m-1} + \delta_{n, m+1} - 2i \Gamma_n \delta_{n,m}.
 \label{A5}
 \ee
The $\delta$-symbol in \eqref{A5} is $2N$-periodic:
\[
\delta_{n,m}= \left \{ \begin{array}{ll}
1,  & n=m \ \mathrm{mod} (2N); \\
0, & \mbox{otherwise}.
\end{array}
\right.
\]

The eigenvalues of $\mathcal L$ were determined in \cite{Liam}:
\begin{align}
\beta=\pm \beta_\alpha, \quad
\beta_\alpha =  2 \sqrt{ \cos^2\left( \frac{k_\alpha}{2} \right) -\gamma^2}>0, 
\nonumber
\\
k_\alpha= \frac{2 \pi}{N} \alpha, \quad \alpha=1,2, ..., N. 
\label{A6b}
\end{align}
The eigenvalues are all real if $\gamma \leq \gamma_c$, where
\be
\gamma_c(N)= \left\{ 
\begin{array}{ll}
0, &    N=\mbox{even}; \\
\sin \left( \frac{\pi}{2N} \right), & N=\mbox{odd}.
\end{array}  \right.
\label{threshold}
\ee
Note that in the necklace with even $N$, the eigenvalues become complex as soon as $\gamma$ is nonzero.
For this reason we are only considering odd $N$ in what follows. 
We are  also assuming that the symmetry is not broken, that is, $\gamma \leq \gamma_c$.

Two eigenvalues, $\beta_N$ and $-\beta_N$, are simple (non-repeated). 
The other $\frac{N-1}{2}$ positive and $\frac{N-1}{2}$ negative eigenvalues have algebraic  multiplicity 2.
Indeed, $\beta_\alpha$ coincides with $\beta_{(N-\alpha)}$ for all $\alpha=1,2, ... N-1$.

Turning to the eigenvectors of $\mathcal L$, one can readily check that 
\be
{\vec v}^{(\alpha)} = \left(e^{ik+i\theta}, e^{ik}, e^{2ik+i \theta}, e^{2ik}, ..., e^{Nik+i\theta}, e^{Nik}\right)^T
\label{A8}
\ee
 is an eigenvector corresponding to  a positive eigenvalue $\beta=\beta_\alpha$.
Here $k=k_\alpha$ is as in \eqref{A6b}  and $\theta=\theta_\alpha$ is defined by
\[
e^{i \theta_\alpha} = \frac{1+ e^{-ik_\alpha}}{2 i \gamma + \beta_\alpha}, 
\quad \alpha=1,2,... N.
\]
It is not difficult to verify that the vectors ${\vec v}^{(\alpha)}$ and ${\vec v}^{(N-\alpha)}$ 
 are linearly independent for all $\alpha=1,2, ..., \frac{N-1}{2}$
 and so each positive eigenvalue  $\beta= \beta_\alpha$  has a geometric multiplicity 2.

The vector 
\be
{\vec w}^{(\alpha)} = \left(e^{ik+ i \phi}, e^{ik}, e^{2ik+i \phi}, e^{2ik}, ..., e^{Nik+i \phi}, e^{Nik}\right)^T,
\label{A50}
\ee
where $k=k_\alpha$ is as in \eqref{A6b}  and $\phi=\phi_\alpha$ is defined by
\[
e^{i \phi_\alpha} = \frac{1+ e^{-ik_\alpha}}{2 i \gamma - \beta_\alpha}, 
\quad \alpha=1,2,... N,
\]
is an eigenvector associated with a {\it negative\/}  eigenvalue $\beta=-\beta_\alpha$.
Since the eigenvectors ${\vec w}^{(\alpha)}$ and ${\vec w}^{(N-\alpha)}$
pertaining to the equal eigenvalues $ -\beta_{N-\alpha}$ and  $-\beta_\alpha$ 
 are linearly independent for any   $1 \leq \alpha \leq \frac{N-1}{2}$,
we conclude that  each repeated negative eigenvalue of the matrix $\mathcal L$ has  a geometric multiplicity 2  as well.

\section{Nonlinear selection rule}
\label{rules}

Returning to the nonlinear dispersive system  \eqref{A1},   we introduce a hierarchy of stretched coordinates $Z_\ell = \epsilon^\ell z$
and 
time scales
$T_\ell =\epsilon^\ell \tau$;
$\ell= 0, 1,2, ...$. In the limit $\epsilon \to 0$ all these variables become independent
and the chain rule gives 
\[
\frac{\partial}{\partial z}= D_0+ \epsilon^2 D_2+   \epsilon^4 D_4 + ..., 
\quad
\frac{\partial}{\partial \tau}= \partial_0+ \epsilon \partial_1+ \epsilon^2 \partial_2+..., 
\]
where $D_\ell=\partial/ \partial Z_\ell$ and 
$\partial_\ell= \partial/ \partial T_\ell$.
Symmetry considerations suggest that the complex modes $\psi_n$ should not depend on the odd coordinates $Z_1, Z_3, ...$, ---  this is why we have omitted the odd terms in the expansion of $\partial_z$.
Expanding 
\[
\psi_n= \epsilon A_n + \epsilon^3 B_n+ ...
\]
and substituting into  \eqref{A1} we equate coefficients of like powers of $\epsilon$. 

The order $\epsilon^1$ gives
\be
i D_0 {\vec A} + \mathcal L {\vec A}=0,
\label{A16}
\ee
where ${\vec A}=(A_1, A_2, ..., A_{2N})^T$
and we have assumed that $\vec A$ does not change on the fast time scale, $T_0$. 
The general solution of \eqref{A16} is given by a linear combination
\be 
{\vec A} = \sum_{\alpha=1}^{N} p^{(\alpha)} {\vec v}^{(\alpha)} e^{i \beta_\alpha z}+  \sum_{\alpha=1}^{N} r^{(\alpha)} {\vec w}^{(\alpha)} e^{-i \beta_\alpha z},
\label{A17}
\ee
where the constant vectors ${\vec v}^{(\alpha)}$ and ${\vec w}^{(\alpha)}$ are as in \eqref{A8} and \eqref{A50}
while
 the scalar coefficients $p^{(\alpha)}$ and $r^{(\alpha)}$ are assumed to
depend on the ``slow" variables $Z_2, Z_4, ...$ and $T_1, T_2, ...$.
The individual terms in \eqref{A17} are commonly referred to as supermodes.
The sum
\eqref{A17} with a specific choice of  coefficients will be called 
a ``linear pattern" in what follows.

To determine nonlinear constraints that select particular linear patterns in the necklace, we 
proceed to the 
 order $\epsilon^3$ which gives a nonhomogeneous system of equations for coefficients $B_n$:
 \be
 i D_0 {\vec B} + \mathcal L {\vec B} = {\vec {\mathcal R}}, 
 \label{A18}
 \ee
 where
 \[
{\mathcal R}_n= - \left( iD_2+\partial_1^2 + 2|A_n|^2 \right) A_n;
 \]
 $n=1,2,..., 2N$. 
The vector function   ${\vec{\mathcal R}}$  will generally have terms that 
are in resonance with the ``frequencies"  of the linear nondispersive system.
The unbounded growth of the coefficients $B_n$ as $z \to \infty$  (and the resulting breakdown of the asymptotic
expansion) can only be avoided  if ${\vec {\mathcal R}}$ is orthogonal to the eigenvectors of the matrix
$\mathcal L$. These orthogonality relations: (a) select the linear patterns that persist in the nonlinear dispersive regime
when the amplitudes of the complex modes are no longer small and the beams are no longer stationary; 
(b)  determine the longitudinal structure 
and temporal evolution of  nonlinear pulses of light.

In the subsequent sections we go over  several possible choices in    \eqref{A17}.

\section{Simultaneous pulses in $2N$ guides} 
\label{sims}

Circular-symmetric distributions of power $|\psi_n|^2$ result by keeping only one supermode
in the linear pattern \eqref{A17}. 
Choosing
\be
{\vec A}= p {\vec v}^{(N)} e^{i \beta z},
\label{A19}
\ee
where 
$\beta=\beta_N$ and $p=p^{(N)}$,
a bounded solution to equations \eqref{A18} (if exists) will have the form
\be
{\vec B}= {\vec{ \mathcal X}} e^{i \beta z},
\label{A20}
\ee
where ${\vec{\mathcal X}}$ satisfies
\be
( \mathcal L -  \beta I)  \vec{ \mathcal X}
=- ( iD_2 p + \partial_1^2 p +   2 |p|^2 p) \vec{v}^{(N)}.
\label{A21}
\ee

The singular system \eqref{A21} admits a  solution 
 if and only
if its right-hand side is orthogonal to the  eigenvector ${\vec v}^{(N)}$
in the sense of the dot product
\be
{\vec y} \cdot {\vec z}= \sum_{n=1}^{2N} y_n z_n.
\label{DP} \ee
[In     equation \eqref{DP},  $\vec{y}$ and $\vec{z}$ are  vectors with complex components.]
Making use of the identity 
\be
{\vec v}^{\, (\alpha)} \cdot {\vec v}^{\, (\alpha)} = (1+ e^{i \theta_N}) N\delta_{\alpha, N}
\label{A14}
\ee
 with $\alpha=N$,  the solvability condition reduces to the nonlinear Schr\"odinger equation
\be  
i\frac{\partial p}{\partial Z_2} + \frac{\partial^2 p}{\partial T_1^2}  +  2 |p|^2 p=0.
\label{A22}
\ee

A localised solution of equation \eqref{A22} is the soliton
\be
p=  e^{i Z_2} \mathrm{sech}  \, T_1,
\label{p1}
\ee
where the amplitude has been set equal to 1. 
(There is no loss in generality in setting the amplitude to unity as it only appears as a coefficient in front of $\epsilon$ when the solution is expressed in the original coordinates.)
The vector function \eqref{A19} with $p$ as in \eqref{p1} 
 describes identical light pulses
 travelling  in $2N$ waveguides
 level with each other. 
All waveguides shine in unison
and with the same intensity: $|\psi_n|^2= \epsilon^2  \mathrm{sech}^2 (\epsilon \tau)$.

Another simultaneous ring of pulses results by 
letting
 $r^{(\alpha)}=0$ for all $\alpha=1,...,N$,
and
 $p^{(\alpha)}=0$ for all $\alpha$ except one particular value $\alpha=\alpha_0$
and its symmetric partner  $\alpha=N-\alpha_0$.  Here $1 \leq \alpha_0 \leq \frac{N-1}{2}$.
Denoting 
\[
p^{(\alpha_0)} \equiv p,  \quad
p^{(N-\alpha_0)} \equiv q, \quad
{\vec v}^{(\alpha_0)} \equiv {\vec v}, \quad
{\vec v}^{(N-\alpha_0)} \equiv {\vec u},
\]
and $\beta_{\alpha_0}=\beta$,
 the linearised pattern \eqref{A17} becomes
\be
{\vec A}= \left( p {\vec v} + q {\vec u}  \right) e^{i \beta z}.
\label{td} 
\ee
A bounded third-order correction $B_n$ has the form \eqref{A20}, 
where the vector $\vec{\mathcal X}$ satisfies the system
\begin{align}
\sum_{m=1}^{2N} \mathcal L_{nm}  \mathcal X_m
- \beta \mathcal X_n   
= -Fv_n  -G u_n      \nonumber \\
-2 \left[
q^2 p^* u_n^2 v_n^* + p^2q^* v_n^2 u_n^*\right]
\label{A31}
\end{align}
with the coefficient functions
 \begin{align} 
F(Z_2,..., T_1, ...) = \left(  iD_2 +\partial_1^2  + 2 |p|^2+ 4|q|^2 \right) p,   \label{F} \\
G(Z_2, ..., T_1, ...) = \left(  iD_2 +\partial_1^2 
        + 4 |p|^2        + 2  |q|^2\right) q.   \label{G}
\end{align}

Since the zero eigenvalue of the matrix $\mathcal L -\beta I$ in the left-hand side of \eqref{A31} has 
 geometric multiplicity 2, 
the nonhomogeneous system \eqref{A31} has two solvability conditions.
Taking
 the scalar product of its right-hand side with $\vec u$ and $\vec v$ produces a pair of amplitude equations:
 \begin{subequations}
 \label{2NLS}
 \begin{align}
i \frac{\partial  p}{\partial Z_2}  + \frac{\partial^2 p}{\partial T_1^2} +   2 ( |p|^2 + 2|q|^2) p=0,
\label{A27} \\
i \frac{\partial  q}{\partial Z_2}  + \frac{\partial^2 q}{\partial T_1^2} +  2 (|q|^2+ 2|p|^2) q=0.
 \label{A28}
\end{align}
\end{subequations}
In obtaining the system \eqref{2NLS}, we  used the following two  identities in addition to the identity  \eqref{A14}:
\begin{align}
 {\vec v}^{(\alpha)} \cdot {\vec v}^{(N-\alpha)}=
 \left( e^{i(k_\alpha+ 2 \theta_\alpha)}+1 \right) N,     \label{Q10}  \\
\sum_{n=1}^{2N} v_n^3 u_n^*= 
\sum_{n=1}^{2N} u_n^3 v_n^*=0.   \nonumber
\end{align}

The power distribution 
associated with a repeated eigenvalue $\beta_{\alpha_0}$  is $z$-independent but not uniform across the necklace.
 Letting, for simplicity,   $p=q$,  equation \eqref{td} gives 
\[
|A_n|^2= 2 |p|^2 \left[1 + \cos (n k_{\alpha_0}) \right].
\]

A localised pattern arises when the soliton solution of \eqref{2NLS} is chosen:
\be
p=q=  \frac{1}{\sqrt 3}         e^{i  Z_2}                  \mathrm{sech} \, T _1.
\label{pqs}
\ee
The vector  \eqref{td} with $p$ and $q$ as in \eqref{pqs} describes a ring-shaped constellation of light pulses
travelling abreast in $2N$ fibers. The pulse power undergoes a sinusoidal variation along the ring.

Earlier studies of 
simultaneous pulses in circular arrays of coupled hermitian waveguides were reported in Refs \cite{syncsol,Akhmediev,Liam}.
In \cite{Akhmediev},  rings of  solitonic pulses with varying power were described as bifurcations of the 
uniformly powered ring. Our perspective here is different; we have considered 
simultaneous pulses as nonlinear perturbations of nonuniform linear patterns.

\section{Uniform breathers}
\label{breathers}

Keeping terms with both positive and negative propagation constants in the linear pattern \eqref{A17}
gives rise to $z$-dependent power distributions. 
The simplest possibility corresponds to retaining just two terms:
\be
{\vec A}= p {\vec v} e^{i \beta z} + q {\vec w} e^{-i\beta z}.
\label{A23}
\ee
Here $\beta=\beta_N$ is a simple positive eigenvalue, while
\[
{\vec v}={\vec v}^{(N)}, \quad
{\vec w}={\vec w}^{(N)}
\]
 are the eigenvectors corresponding to $\beta_N$ and 
its negative, respectively. 
With this choice,
the bounded solution of  equation \eqref{A18} is 
\be
{\vec B}= \vec{\mathcal X} e^{i \beta z} + \vec{\mathcal Y} e^{-i \beta z} +\vec{\mathcal M}  e^{3i \beta z} + \vec{\mathcal N} e^{-3i \beta z},
\label{A24}
\ee
where the amplitudes $\vec{\mathcal X}$  and $\vec{\mathcal Y}$ satisfy nonhomogeneous algebraic equations with singular matrices:
\begin{align}
   \left(  \mathcal L - \beta I \right) \vec{\mathcal X} 
=- F \vec{v}, 
\label{A25}  \\
\left(  \mathcal L +  \beta I \right) \vec{\mathcal Y}
=- G \vec{w}.
\label{A26}
\end{align}
Here $F$ and $G$ are as in \eqref{F} and \eqref{G}.

Equation \eqref{A25} admits a solution if and only if its right-hand side is orthogonal to ${\vec v}$ while
the right-hand side of  \eqref{A26}  should be orthogonal to ${\vec w}$. (Here orthogonality is understood in the sense of the 
dot product \eqref{DP}.)
Using \eqref{A14} and the identity
\[
 {\vec w}^{\, (\alpha)} \cdot {\vec w}^{\, (\alpha)} = (1+ e^{i \phi_N}) N\delta_{\alpha, N}
\]
 with $\alpha=N$, these orthogonality constraints translate into equations 
\eqref{2NLS}.

Letting
$q=p$, equation \eqref{A23} gives rise to an oscillatory power distribution:
\begin{align*}
 |A_{2m-1}|^2= 4 |p|^2 \sin^2 (\beta z + \theta_N),
\nonumber \\
|A_{2m}|^2= 4 |p|^2 \cos^2 (\beta z), 
\end{align*} 
$m=1,2, ..., N$.
This describes a flashing necklace: all  odd waveguides blink in unison
and all even waveguides reach their maximum power at the same $z$, but there is a lag between the odd and even. 
Note that the flashing is uniform: the maximum power is the same for all waveguides.

The soliton solution \eqref{pqs} 
of the 
 system \eqref{2NLS}
provides an envelope for  a finite-duration sequence of  short  flashes
 in the necklace --- a spatio-temporal pattern commonly referred to  as a  breather.
 Breathers in fiber directional couplers (that is, in  necklaces consisting just of 2 waveguides, with no gain or loss)
 were described numerically and variationally \cite{coupler}.  
 For the asymptotic descriptions and nonhermitian extensions, see \cite{BSSDK}.

\section{Nonuniform flashing} 
\label{Nonuni}

A set of slightly more complex patterns results by letting,
in equation \eqref{A17},
$r^{(\alpha)}=0$ 
and $p^{(\alpha)}=0$ for all $\alpha=1,2,..., N$ except  one particular value  $\alpha=\alpha_0$
($1 \leq \alpha_0 \leq \frac{N-1}{2}$)
and its symmetric partner  
$\alpha= N-\alpha_0$. 
Denoting 
\[
p^{(\alpha_0)} \equiv p_1,   \  r^{(\alpha_0)} \equiv q_1,   \   r^{(N-\alpha_0)} \equiv p_2,   \
p^{(N-\alpha_0)} \equiv q_2
\]
and 
$\beta_{\alpha_0}=\beta$,
 equation \eqref{A17} becomes
 \begin{subequations}
 \label{Y}
\be
{\vec A}= {\vec \eta}  e^{i \beta z} + {\vec \xi} e^{-i \beta z} 
\ee
where
\be
{\vec \eta}= p_1 {\vec v}^{(\alpha_0)}     + q_2 {\vec v}^{(N-\alpha_0)}, 
\
{\vec \xi}= q_1   {\vec w}^{(\alpha_0)}          + p_2         {\vec w}^{(N-\alpha_0)}.         
\ee
\end{subequations}

The  next-order correction has the form  \eqref{A24}
where $\vec{\mathcal X}$ and $\vec{\mathcal Y}$ satisfy
\begin{align}
(\mathcal L - \beta I) {\vec {\mathcal X}}= - {\vec {\mathcal F}},              \label{A40}               \\
(\mathcal L + \beta I) {\vec {\mathcal Y}}= - {\vec {\mathcal G}},   \label{A41} 
\end{align}
with
\begin{align*}
\mathcal F_n=   iD_2  \eta_n +\partial_1^2  \eta_n    +   2( |\eta_n|^2+ 2 |\xi_n|^2)   \eta_n,  \\
 \mathcal G_n=   iD_2 \xi_n +\partial_1^2  \xi_n    +   2(2 |\eta_n|^2+  |\xi_n|^2) \xi_n.
  \end{align*}

The zero eigenvalue of the matrix $\mathcal L -\beta I$ in  equation \eqref{A40} has geometric multiplicity 2,
and the same is true for the zero eigenvalue of the matrix $\mathcal L + \beta I$ in \eqref{A41}.
Evaluating the dot product of the right-hand side of \eqref{A40} with the vectors ${\vec v}^{(\alpha_0)} $  and ${\vec v}^{(N-\alpha_0)} $, and 
then taking the product of the right-hand side 
of \eqref{A41} with  ${\vec w}^{(\alpha_0)}$    and ${\vec w}^{(N-\alpha_0)}$,
we arrive at a system of four amplitude equations:
\begin{subequations}
\label{4NLS}
\begin{align}
i   \frac{\partial  p_\mu}{\partial Z_2}  +\frac{\partial^2  p_\mu}{\partial T_1^2}  +2\left(|p_\mu|^2+2 |p_{\mu+1}|^2  \right) p_\mu
  \nonumber \\
+4 \left(|q_1|^2+ |q_2|^2\right) p_\mu
+ 4 q_1q_2 p_{\mu+1}^*=0,
\label{D05}\\
i   \frac{\partial  q_\mu}{\partial Z_2}  + \frac{\partial^2  q_\mu }{\partial T_1^2} +2\left(|q_\mu|^2+ 2|q_{\mu+1}|^2  \right)q_\mu
  \nonumber \\
+4\left( |p_1|^2+ |p_2|^2\right) q_\mu
+ 4 p_1p_2 q_{\mu+1}^*=0,
\label{D06}
\end{align}
\end{subequations}
where $\mu=1,2$. 
In \eqref{4NLS},  we use the cyclic notation for the indices: 
$p_{3}$ should be understood as $p_1$ and $q_{3}$ as $q_1$.

The system \eqref{4NLS}   is invariant under a 3-parameter 
transformation 
\begin{subequations}\label{inv}
\be
q_\mu \to e^{i \varphi_\mu} q_\mu, \quad p_\mu \to e^{i \vartheta_\mu} p_\mu,
\quad (\mu=1,2),
\ee
where $\varphi_{1,2}$ and $\vartheta_{1,2}$ are four constant angles satisfying
\be
\varphi_1+ \varphi_2 =\vartheta_1 + \vartheta_2.
\label{cosnt} \ee
\end{subequations}
Solutions that are related by the transformation \eqref{inv} will be regarded equivalent.

It is convenient to introduce vector notation for the four-component columns:
\[
\mathbf{ \Phi} =\left( \begin{array}{r}p_1 \\ p_2 \\ q_1   \\ q_2 
\end{array}
\right).
\]
There are ${ \textstyle \left( \begin{array}{c} 4 \\ 2 \end{array} \right)} =6$ nonequivalent soliton solutions with two nonzero components:
\begin{align*} 
\mathbf{\Phi}^{(12)}=
\left( \begin{array}{r} 1 \\ 0 \\ 0 \\ 1
\end{array}
\right)f,  \quad
\mathbf{\Phi}^{(21)}=
\left( \begin{array}{r} 0 \\ 1 \\ 1 \\ 0
\end{array}
\right)f,  \quad
 \mathbf{\Phi}^{(11)}=
\left( \begin{array}{r} 1\\ 0 \\ 1 \\ 0
\end{array}
\right)f,   \nonumber \\
\mathbf{\Phi}^{(22)}=
\left( \begin{array}{r} 0 \\ 1 \\ 0 \\ 1
\end{array}
\right)f, \quad
\mathbf{\Phi}^{(p)}=
\left( \begin{array}{r} 1 \\ 1 \\ 0 \\ 0
\end{array}
\right)f, 
\quad
\mathbf{\Phi}^{(q)}=
\left( \begin{array}{r} 0 \\ 0 \\ 1 \\ 1
\end{array}
\right)f,
\end{align*}
where  $f$ accounts for the large-scale space-time variation  of the pattern:
\be
f(Z_2,T_1)=  \frac{1}{\sqrt 3}         e^{i  Z_2}                  \mathrm{sech} \,  T_1.
\label{Y3}
\ee

The solution $\mathbf{\Phi}^{(12)}$   reproduces  equation
\eqref{td} with $p$ and $q$ as in \eqref{pqs}. This solution as well as 
 $\mathbf{\Phi}^{(21)}$ describe  constellations of   $2N$
 pulses travelling abreast, with their power varying along the necklace.
On the other hand, 
 $\mathbf{\Phi}^{(11)}$   and  $\mathbf{\Phi}^{(22)}$ define uniformly     flashing 
 patterns similar to \eqref{A23}.

Deferring the intepretation of  $\mathbf{\Phi}^{(p)}$ and  $\mathbf{\Phi}^{(q)}$ to the next section, 
here we consider two more soliton solutions of  the system \eqref{4NLS}.
Both solutions have all their components nonzero:
%
\be
\mathbf{\Phi}^{(A)} =
\frac{1}{\sqrt 3}
\left( \begin{array}{r} 1 \\ 1\\ 1\\ 1
\end{array}
\right)f,
 \quad
  \mathbf{\Phi}^{(B)}= \sqrt{\frac35}
\left( \begin{array}{r} 1 \\ 1 \\ 1 \\ -1
\end{array}
\right)f,
\label{Phimp}
\ee
where
$f(Z_2,T_1)$ is as in \eqref{Y3}.

The power load of individual waveguides associated with the solution $\mathbf{\Phi}^{(A)}$ is given by
\begin{align}
|A_{2m-1}|^2= \frac{16}{3}  |f|^2 \cos^2 \left( \frac{2m-1}{2}k\right)    \nonumber \\     \times   \sin^2 \left( \beta z  + \theta           +\frac{k}{2}       \right),   \nonumber \\
|A_{2m}|^2=\frac{16}{3} |f|^2 \cos^2 (mk) \cos^2 (\beta z),  \label{AA2m}
\end{align}
while the soliton $\mathbf{\Phi}^{(B)}$ 
carries the following power distribution:
 \begin{align}
|A_{2m-1}|^2=    \frac{12}{5} |f|^2+   \frac{12}{5} |f|^2
\sin\left[ (2m-1)k\right]     \nonumber \\
      \times        \sin (2 \beta z + 2 \theta +k),       \nonumber \\
|A_{2m}|^2=  \frac{12}{5}   |f|^2 - \frac{12}{5}   |f|^2 \sin(2mk) \sin (2 \beta z).
\label{AA2p}
\end{align}
In either of these equations,  $m=1,2, ...N$, and $\beta=\beta_{\alpha_0}$, $k=k_{\alpha_0}$, $\theta=\theta_{\alpha_0}$.
Both  \eqref{AA2m}    and  \eqref{AA2p}
 represent flashing patterns, or breathers, where all odd and all even waveguides flash synchronously.
 The maximum power attainable  in an individual waveguide
 undergoes  a sinusoidal variation along the necklace.

\section{Gyrating solitons} 
\label{Gyro1} 
\subsection{Single-frequency pattern}
\label{jivers}

\begin{figure}
\begin{center}
 \includegraphics[width=\linewidth]{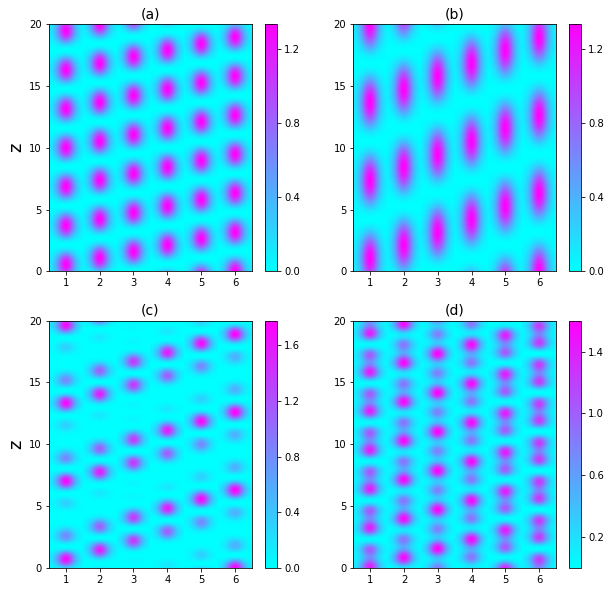}\end{center}
\caption{Four types of gyrating solitons in a necklace of $2N=6$ waveguides.
 Each panel consists of six vertical lanes displaying the $(\tau,z)$-distribution of optical power in six waveguides.
 The horizontal side of each lane represents 
  a short period of time,  $-20<\tau <20$; the $\tau$-axis is not marked or labelled. The vertical coordinate measures the length along the waveguides, with $0 \leq z \leq 20$.
(a)  jiver soliton \eqref{gyro1}.  The panel shows the power distribution \eqref{spr}.
(b) the power density \eqref{Q12} corresponding to the waltzer \eqref{gyro2}. (c):  power distribution       \eqref{Ami}      
associated with the multiflash gyrator $A$.
(d): power pattern  \eqref{Api} of the multiflash solution
 $B$.
 In all panels, $\gamma= 0$ and  $\alpha=1$. All solitons have the inverse-width  parameter $\epsilon =0.1$. 
 \label{four}}
 \end{figure}

The solitons
$\mathbf{\Phi}^{(p)}$ and  $\mathbf{\Phi}^{(q)}$
represent
light pulses gyrating around the necklace.

The power distribution associated with $\mathbf{\Phi}^{(p)}$ has the form of a spiral wave  (Fig \ref{four}(a)):
\begin{align}
|A_{2m-1}|^2= 4 |f|^2 \sin^2 (km+\beta z+\theta), \nonumber  \\
|A_{2m}|^2= 4 |f|^2 \cos^2 (km +\beta z).
\label{spr}
\end{align}
Here  $m=1,2,...N$
and the parameters are
$k= k_{\alpha}$, $\beta= \beta_{\alpha}$ and $\theta=\theta_{\alpha}$.
To simplify the notation,  we have dropped the subscript 0 from  the index $\alpha$  ($1 \leq \alpha \leq \frac{N-1}{2}$).

To establish whether the soliton is gyrating clockwise or counter-clockwise, 
we need to determine which of the two neighbours of
the $2m$-th waveguide will flash immediately after  the $2m$-th guide has.
Assume that  the $2m$-th waveguide attains its maximum power at the point $z=z_0$.
Then the closest maximum of $|A_{2m+1}|^2$ to the right of $z_0$ is at $z=z_0+ \Delta_{2m+1}$, and the nearest maximum of $|A_{2m-1}|^2$ to the right of $z_0$ 
is at $z= z_0+ \Delta_{2m-1}$, where the delay intervals are given by
\be
 \Delta_{2m+1}= \frac{1}{\beta}  \left( \frac{\pi}{2} -\theta -k \right)   \label{2MM}
\ee
and
\be
 \Delta_{2m-1}= \left\{
\begin{array}{rc}
- \frac{1}{\beta}  \left( \frac{\pi}{2}+ \theta \right) &  \mbox{if} \  \theta< - \frac{\pi}{2}, \\
\frac{1}{\beta}  \left( \frac{\pi}{2} - \theta \right) &  \mbox{if} \ \theta >-\frac{\pi}{2}.
\end{array} \right.  \label{2MP}
\ee

Comparing the lags \eqref{2MM} and \eqref{2MP} one can readily check that 
the $(2m+1)$-th guide flashes sooner respectively later than the $(2m-1)$-th one  if $\gamma < \gamma_\alpha$
respectively $\gamma > \gamma_\alpha$,
where 
\[
\gamma_\alpha= \cos^2 \frac{k_\alpha}{2}.
\]

Let
\be
\alpha_c(N)=  \mathrm{floor}   \left[ \frac{N}{\pi} \arccos  \left(  \gamma_c^{1/2}\right)\right],
\label{floor}
\ee
where
$\mathrm{floor}[x]$ stands for the greatest integer less than or equal to $x$, while $\gamma_c= \sin \left( \frac{\pi}{2N} \right)$
 is the linear $\mathcal{PT}$-symmetry breaking threshold  given by equation \eqref{threshold}.
For all $\alpha \leq \alpha_c$
we have $\gamma_\alpha \geq \gamma_c$. Since we are considering a necklace  operating in the stable regime $(\gamma < \gamma_c)$, then,
  assuming that the waveguides are numbered against the clock,
we conclude that 
the soliton $\mathbf{\Phi}^{(p)}$ with any $\alpha=1,2, ..., \alpha_c$  and regardless of $\gamma$,  is gyrating counterclockwise.

\begin{figure}
\begin{center}
 \includegraphics[width=\linewidth]{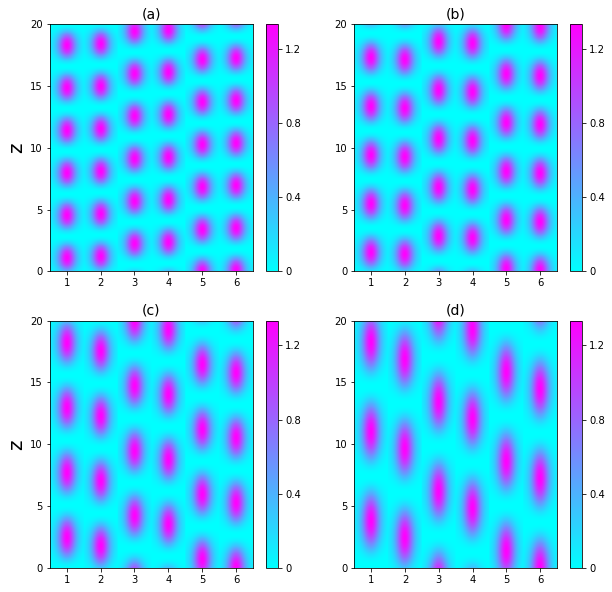}\end{center}
\caption{The transition 
  from counterclockwise to clockwise gyration in the necklace of 6 waveguides. 
    All four panels pertain to the  same jiver soliton  as in Fig \ref{four}(a) 
but with different $\gamma$.
  As in Fig \ref{four}, each panel consists of 6 vertical lanes. The $n$-th lane
shows $|A_n(\tau,z)|^2$,  the power density in the $n$-th waveguide.
 The horizontal side of each  lane represents 
  a short period of time,  $-20<\tau <20$, with the $\tau$-axis  not marked or labelled. The vertical coordinate measures the length along the waveguides.   
 All four power distributions are given by equation \eqref{spr} where $\alpha=1$, $\epsilon=0.1$ while $\gamma$ varies:
  (a)
 $\gamma=0.20$;         (b)        $\gamma=0.30$;        (c)      $\gamma=0.40$;          (d)     $\gamma=0.45$.
 The transition occurs as $\gamma$ is raised through $\gamma_1= 0.25$.
 \label{reverse}}
 \end{figure}

By contrast, the sense of gyration of the soliton $\mathbf{\Phi}^{(p)}$ with $\alpha= \alpha_c+1, ...,   \frac{N-1}{2}$ does depend on $\gamma$. 
The corresponding transition values $\gamma_\alpha$ lie under the $\mathcal{PT}$-symmetry breaking threshold.
When $0 \leq \gamma < \gamma_\alpha$, the soliton gyrates counterclockwise but when $\gamma_\alpha < \gamma < \gamma_c$, 
it revolves in the clockwise direction. This crossover is illustrated by Fig \ref{reverse}.

The behaviour of the solitons $\mathbf{\Phi}^{(q)}$ is opposite to that of $\mathbf{\Phi}^{(p)}$. Namely,  pulses
with  $\alpha=1,2, ..., \alpha_c$ are gyrating clockwise for all $\gamma$.  Those with  $\alpha=  \alpha_c+1, ...,  \frac{N-1}{2}$
are also revolving clockwise for small $\gamma$ but their
  direction of gyration
can be reversed by raising $\gamma$ above  $\gamma_\alpha$.

 The two gyrating solitons whose linear patterns  are given by    equation  \eqref{Y} with the coefficients defined by the vector $\mathbf{\Phi}^{(p)}$ or $\mathbf{\Phi}^{(q)}$,
 can be written in a unified way as
  \be
 \label{gyro1}
 {\vec \psi} =     \epsilon      \frac{
  {\vec v}^{(\alpha)} e^{i \beta_\alpha z} + {\vec w}^{(N-\alpha)} e^{-i \beta_\alpha z}}{\sqrt 3} 
          e^{i \epsilon^2 z}                  \mathrm{sech} (\epsilon \tau)
 + O(\epsilon^3), 
 \ee
 where $1 \leq \alpha \leq N-1$. 
 Solitons with $\alpha= N-\alpha_c, ..., N-1$ are gyrating clockwise and those with $\alpha=1, ..., \alpha_c$ are moving against the clock. For $\alpha= \alpha_c+1, ...,  N-\alpha_c-1$, the direction of
 gyration
 is controlled by the choice of $\gamma$.

Before turning to other types of gyrating pulses we note two more characteristics of the solitons \eqref{gyro1} that 
can be controlled in the nonhermitian situation. 
Namely, by varying the gain-loss coefficient one can change the length of the pulse and its
period of revolution around the necklace. Both of these quantities are given by the $z$-period of the power density \eqref{spr}.
The length of two particular pulses with $\alpha= \frac{N \pm 1}{2}$ can even be sent to infinity --- one just needs to tune $\gamma$  to $\gamma_c$.
(The reason is that the propagation constant $\beta_{(N \pm 1)/2}  \to 0$ 
 as $\gamma \to \gamma_c$.)

 Fig \ref{reverse}  exemplifies the  change in flash duration with a sequence of 
   four values of $\gamma$ from  the interval $(0, \gamma_c)$.

\subsection{Two-frequency pattern}

A quasiperiodic pattern that does not fit into the general Ansatz \eqref{Y}
 combines eigenvectors associated with a repeated and a single eigenvalue:
\be
\vec A = p {\vec v}^{(\alpha)} e^{i \beta_\alpha z} + q {\vec v}^{(N)} e^{i \beta_N z}.
\label{Q11}
\ee
Here $\alpha$ is an arbitrarily chosen mode number,
 $1 \leq \alpha \leq N-1$. With this choice, the right-hand side of equation \eqref{A18} 
features two resonant terms proportional to $e^{i \beta_\alpha z}$ and $e^{i \beta_N z}$, respectively.
Since $\beta_\alpha$ is a repeated eigenvalue, the former term imposes two solvability conditions.
With the help of \eqref{A14}, we verify that  one of these is trivially satisfied. The other solvability condition, together with the solvability constraint
associated with the propagation constant $\beta_N$, comprise the system \eqref{2NLS}. (The derivation makes use of the identities
\eqref{A14} and \eqref{Q10}.)

Like  the distribution \eqref{spr} before, 
the power density associated with the pattern \eqref{Q11} has the form of a spiral:
\begin{align}
|A_{2m-1}|^2= 4 |p|^2 \cos^2 \left[ \frac{  k_\alpha m+(\beta_\alpha-\beta_N)z+ \theta_\alpha-\theta_N }{2}\right],         \nonumber  \\
|A_{2m}|^2= 4 |p|^2 \cos^2  \left[ \frac{  k_\alpha m+(\beta_\alpha-\beta_N)z }{2}\right],   
\label{Q12}
\end{align}
where $m=1,2, ..., N$ and we have assumed a simple reduction of the system \eqref{2NLS}: $p=q$.
(See Fig \ref{four} (b).)
A localised pattern corresponds to the soliton solution of that system,
equation  \eqref{pqs}.

The self-contained form of the solution whose linear order is given by equation
\eqref{Q11} with $p$ and $q$ as in \eqref{pqs},  is
  \be
 \label{gyro2}
 {\vec \psi} =     \epsilon      \frac{
  {\vec v}^{(\alpha)} e^{i \beta_\alpha z} + {\vec v}^{(N)} e^{i \beta_N z}}{\sqrt 3} 
          e^{i \epsilon^2 z}                  \mathrm{sech} (\epsilon \tau)
 + O(\epsilon^3).
 \ee
This is  a new gyrating soliton in the necklace. 
An argument similar to the one in section \ref{jivers} shows that 
  the solitons with $\alpha \geq \frac{N+1}{2}$ are gyrating clockwise
 while those with $\alpha \leq \frac{N-1}{2}$ are moving against the clock.

The panels (a) and (b) of
Fig \ref{four}  illustrate the difference between the two types of gyrating solitons in the hermitian 
  necklace of $6$ waveguides.
 The spiral pattern \eqref{Q12}  displays a longer period of revolution around the necklace
than the 
 pattern \eqref{spr}.   
By the time the soliton \eqref{gyro2}  completes just one round of its ``waltz" 
around the necklace, its more agile counterpart \eqref{gyro1} will  have ``jived" around twice.
For ease of reference, we dub the gyrating solitons \eqref{gyro1} and \eqref{gyro2} the {\it jiver\/}  and the {\it waltzer}, respectively.

\section{Multiflash gyration}
\label{Gyro2} 

When the waveguides are linear and nondispersive, that is, 
when the necklace is described by the system \eqref{A1} with neither cubic nor time-derivative terms included, 
any set of coefficients $p^{(\alpha)}$ and $r^{(\alpha)}$ 
in \eqref{A17} defines a pattern in the necklace. 
However, only a handful of those patterns persist the addition of nonlinear and dispersive terms to \eqref{A17}.

In this section we identify two more spiral patterns associated with
 gyrating solitons.  
The patterns in question 
 generalise the  two-mode combination \eqref{Q11}. They involve
an eigenvector ${\vec v}^{(\alpha)}$  associated with a repeated eigenvalue $\beta_\alpha$
(where $1 \leq \alpha \leq N-1$),
its mirror-reflected  conterpart ${\vec w}^{(N-\alpha)}$ associated with the negative propagation constant $-\beta_\alpha$,
and the eigenvectors ${\vec v}^{(N)}, {\vec w}^{(N)}$ corresponding to the pair of single eigenvalues $\pm \beta_N$:
\begin{align*}
{\vec A}= p_1 {\vec v}^{(\alpha)} e^{i \beta_\alpha z} +
p_2 {\vec w}^{(N- \alpha)} e^{-i \beta_\alpha z} \nonumber \\
+ q_1 {\vec v}^{(N)} e^{i \beta_N z} +
q_2 {\vec w}^{(N)} e^{-i \beta_N z}.
\end{align*}
 This time, the right-hand side of equation \eqref{A18} 
has four resonant terms proportional to $e^{\pm i \beta_\alpha z}$ and $e^{\pm i \beta_N z}$. Two of the six solvability conditions are 
satisfied automatically while the remaining four amount to the system 
\eqref{4NLS}.

Two nonequivalent solutions 
of the system \eqref{4NLS}  with all components nonzero are given by equations \eqref{Phimp}.
The power distribution associated with the solution
 $\mathbf{\Phi}^{(A)}$ has the form   
 \begin{align}
 |A_{2m-1}|^2= \frac43 |f|^2  \left[ \sin(\beta_N z + \theta_N)  \right.   \nonumber  \\  \left.   + \sin (m k_\alpha + \beta_\alpha z+ \theta_\alpha) \right]^2,
 \nonumber \\
 |A_{2m}|^2 =\frac43 |f|^2 \left[ \cos(\beta_Nz) + \cos(m k_\alpha + \beta_\alpha z)\right]^2.
 \label{Ami}
 \end{align}
 Here  $m=1,2,...N$ and the slowly changing amplitude  $f$ is  given by \eqref{Y3}. 
   The power distribution \eqref{Ami}  describes several flashes of unequal brightness appearing in rapid succession.
 The string of pulses gyrates around the necklace as a whole, with the ordering of bright and dim flashes changing from one waveguide to another.

 Fig \ref{four}(c) illustrates a multiflash string  \eqref{Ami}    in a necklace of $2N=6$ guides.
 In this case the string comprises a bright flash and one or two dim pulses appearing short distances apart.
 In waveguides on one side of the necklace, the bright flash comes before the dim signal  and on the other side the bright pulse 
 follows the dim one.

 The power distribution corresponding to the solution
 $\mathbf{\Phi}^{(B)}$ is
 \begin{align}
 |A_{2m-1}|^2= \frac{12}{5} |f|^2  \left[ \cos^2(\beta_N z + \theta_N)  \right.   \nonumber  \\  \left.   + \sin^2 (m k_\alpha + \beta_\alpha z+ \theta_\alpha) \right],
 \nonumber \\
 |A_{2m}|^2 =\frac{12}{5}  |f|^2 \left[ \sin^2(\beta_N z) + \cos^2(m k_\alpha + \beta_\alpha z)\right].
 \label{Api}
 \end{align}
  Here  $m=1,2,...N$ and the coefficient function $f$ is as in \eqref{Y3}. 
 As the power pattern \eqref{Ami}, the distribution \eqref{Api} describes a multiflash string gyrating around the necklace
 (see Fig \ref{four}(d)).

    Although the multiflash patterns  have more complex power distributions  than the spirals
   \eqref{spr} and \eqref{Q12}, they   play 
 an important role in the dynamics of the necklace.  Numerical simulations indicate that the
 multiflash gyrating strings may emerge as products of the evolution of the
 unstable  single-pulse gyrators \eqref{spr}. (See section \ref{stab} below.)

 For future reference, we reproduce the multiflash gyrating solitons in a self-contained form:
  \begin{align}
{\vec \psi}_A= \epsilon \left(  {\vec v}^{(\alpha)} e^{i \beta_\alpha z} +
 {\vec w}^{(N- \alpha)} e^{-i \beta_\alpha z}  + {\vec v}^{(N)} e^{i \beta_N z}
 \right.  \nonumber \\
 \left. +
 {\vec w}^{(N)} e^{-i \beta_N z}    \right) \frac{e^{ i \epsilon^2 z}}{3}  \mathrm{sech} ( \epsilon \tau)+ O(\epsilon^3);
\label{Q44}
\end{align} 
 \begin{align}
{\vec \psi}_B= \epsilon \left(  {\vec v}^{(\alpha)} e^{i \beta_\alpha z} +
 {\vec w}^{(N- \alpha)} e^{-i \beta_\alpha z}  + {\vec v}^{(N)} e^{i \beta_N z}
 \right.  \nonumber \\
 \left. 
 -
 {\vec w}^{(N)} e^{-i \beta_N z}    \right) \frac{e^{ i \epsilon^2 z}}{\sqrt 5}  \mathrm{sech} ( \epsilon \tau)+ O(\epsilon^3).
\label{Q141}
\end{align} 
As the notation suggests, we call \eqref{Q44} and \eqref{Q141} the $A$- and $B$-multiflash gyrator, respectively.

\section {Soliton dynamics}
\label{Dynamics}
\subsection{Stability and scattering of gyrating solitons} 
\label{stab}

\begin{widetext}

\begin{figure}
\begin{center}
\includegraphics[width=0.49\linewidth]{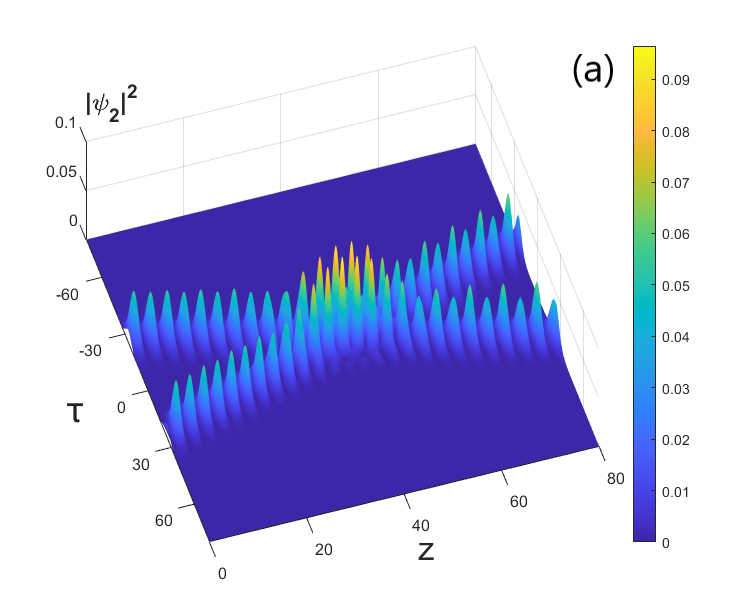} 
\includegraphics[width=0.49\linewidth]{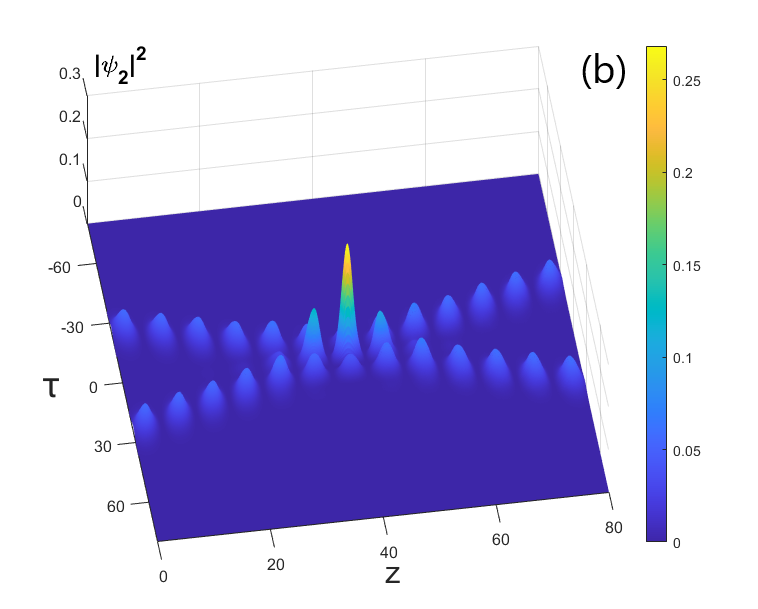} \\
\includegraphics[width=0.49\linewidth]{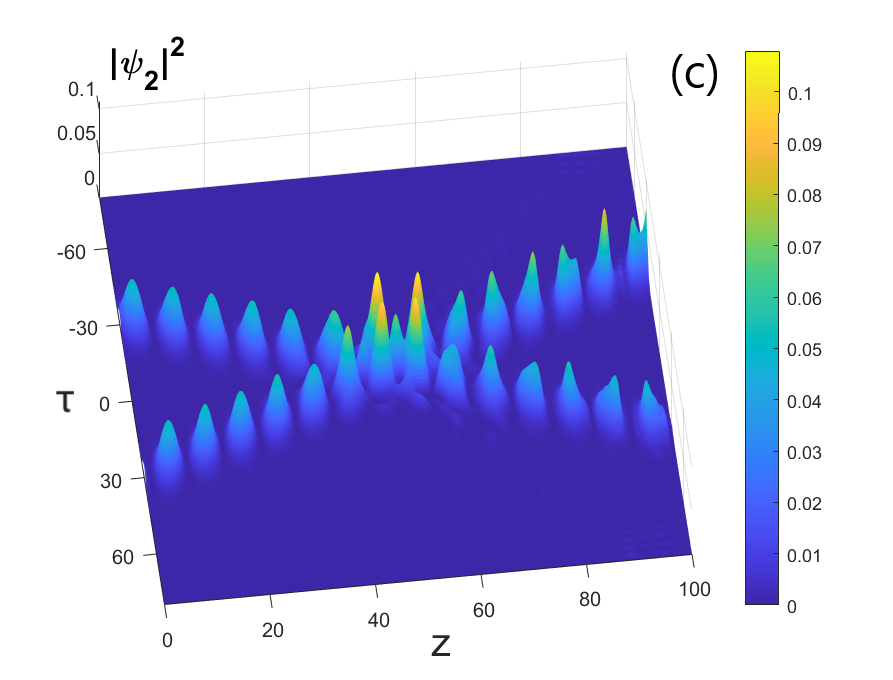} 
\includegraphics[width=0.49\linewidth]{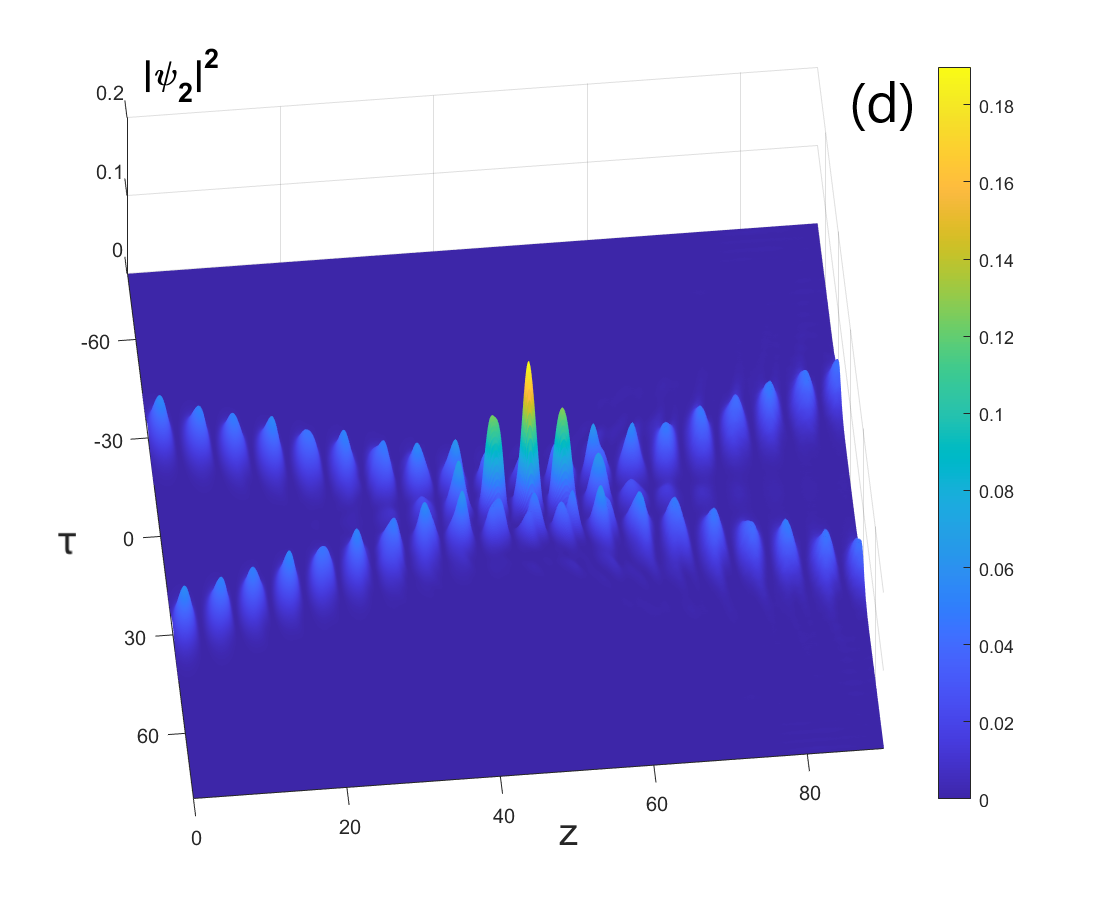} 
\end{center}
\caption{
Scattering of gyrating solitons in the necklace of $2N=6$ waveguides. Shown is $|\psi_2|^2$,
the power density in the second waveguide. 
(a,c) Collision of  two ``jiving" solitons with opposite sense of gyration. The initial condition is \eqref{CA} 
with $\alpha=1$, $\epsilon=0.2$,  and $V=0.6$. Panel (a) corresponds to $\gamma=0$ and panel (c) to  $\gamma=0.45$.
In either case,
the products of collision constitute solitons with the modulated flashing amplitude. (The modulation is
manifested in the alternation of peaks of unequal height.)
(b,d) Collision of two waltzers. The initial condition is \eqref{compo} with $\alpha=1$, $\epsilon=0.2$,  and $V=0.6$. Panels (b) and (d) depict the scattering process in
the system \eqref{A1} with $\gamma=0$ and $\gamma=0.45$, respectively.
In both cases the post-collision solitons restore their original shape.
 \label{collision}}
 \end{figure}
\end{widetext}

The comprehensive stability analysis  of gyrating solitons is beyond the 
scope of the present study. Here, we restrict ourselves to  a few sets of numerical simulations
verifying that these novel objects do not blow up, disperse or transmute into non-gyrating  localised structures within a short period of time.

All our computer simulations were carried out on the necklace of six  waveguides ($N=3$). We considered 
 the system \eqref{A1} both in the hermitian ($\gamma=0$)  and  $\mathcal{PT}$-symmetric
  ($\gamma \neq 0$) situation.

Our first series of simulations involved 
the  ``jiving" soliton, equation  \eqref{gyro1} with $\alpha=1$
(Fig \ref{four} (a)).
The jiver was found to be weakly unstable, both for $\gamma=0$ and $\gamma \neq 0$.
Choosing 
 the initial condition in the form 
\eqref{gyro1}  with  $\epsilon=0.1$ or $\epsilon=0.2$,
and neglecting the $O(\epsilon^3$) terms, 
 the resulting oscillatory solution  was seen to slowly evolve into 
the multiflash solution \eqref{Q44}.
The pattern shown in Fig \ref{four} (a) would gradually transform into the density profile 
of Fig \ref{four} (c).

By contrast,  the ``waltzing" soliton in the same system has turned up to be stable for all values of $\gamma$ that we examined, including $\gamma=0$. 
Random noise  added to the initial condition in the form 
  \eqref{gyro2} with $\alpha=1$ and $\epsilon=0.1$ or $0.2$,
   did not produce any measurable growth of the perturbation.
  The pattern shown in Fig \ref{four} (b) would remain visibly unchanged.

It is instructive to compare  the interaction of two jivers  to  the scattering of two waltzing solitons.
We note that the system \eqref{A1} has the Galilei invariance; namely, if $\psi_n(\tau,z)$ is a solution, then so is
\[
{\tilde \psi}_n (\tau,z)  \equiv  e^{i \frac{V}{2} \left( \tau- \frac{V}{2} z \right) } \psi_n (\tau -Vz,z).
\]
In particular, if $\psi$ is a quiescent, unmoving, soliton, then ${\tilde \psi}$ gives the pulse travelling with the velocity $V$.

Making use of the Galilei transformation  we set up
an initial condition for the collision of  two  clockwise-gyrating  jivers with equal amplitudes and equal oppositely-directed velocities:
\begin{align} 
\psi_n   &  = \frac{\epsilon }{\sqrt 3} \left( v_n^{(\alpha)} + w_n^{(N-\alpha)} \right)   
\left\{ e^{i     \frac{V}{2} \tau}     \mathrm{sech} \left[ \epsilon  (\tau+\tau_0) \right]      \right.  \nonumber    \\     & +
\left.  e^{-i  \frac{V}{2} \tau} \mathrm{sech} \left[ \epsilon  (\tau-\tau_0) \right] \right\}; \quad n=1,..., 2N.
\label{CC}
\end{align}
The collision of a clockwise and anti-clockwise jiving solitons was simulated using an initial condition
 of the form
\begin{align} 
\psi_n & =  \frac{\epsilon }{\sqrt 3}
\left\{   
\left(    v_n^{(\alpha)} + w_n^{(N-\alpha)} \right)    e^{i   \frac{ V}{2} \tau}    \mathrm{sech} [\epsilon  (\tau+\tau_0)]    \right.   \nonumber
\\
   & +   \left.
\left(     v_n^{(N-\alpha)}+ w_n^{(\alpha)} \right) e^{-i   \frac{V}{2} \tau}     \mathrm{sech} [\epsilon  (\tau-\tau_0)]
\right\},
\label{CA}
\end{align}
where $n=1, ..., 2N$.

Despite the jiver's weak instability, both the co-gyrating and counter-gyrating soliton pair emerged from  the collision unscathed.
In the case of either initial condition, equation \eqref{CC} or \eqref{CA}, 
the only effect of interaction was an acquired modulation
of each soliton's oscillation amplitude  (Fig \ref{collision}(a,c)).

Turning to 
 the collision of two waltzers,
 we set the initial condition  in the form 
 \begin{align}
 \label{compo}
 {\vec \psi} =       \frac{  \epsilon  }{\sqrt 3} 
 ( {\vec v}^{(\alpha)}  + {\vec v}^{(N)} )
  \left\{ e^{i     \frac{V}{2} \tau}     \mathrm{sech} \left[ \epsilon  (\tau+\tau_0) \right]      \right.  \nonumber    \\      +
\left.  e^{-i  \frac{V}{2} \tau} \mathrm{sech} \left[ \epsilon  (\tau-\tau_0) \right] \right\}.  
 \end{align}
In this case, the scattering was seen to be elastic.
The solitons would emerge  without any change in the amplitude, velocity or the gyrating pattern
(Fig \ref{collision} (b,d)). \\

\subsection{Vector Schr\"odinger equations}

The two-component amplitude equation \eqref{2NLS} and its four-component counterpart \eqref{4NLS} are
worth commenting upon.

The vector nonlinear Schr\"odinger equation \eqref{2NLS} appeared in a large number of contexts 
and significant wealth of knowledge about its solutions has been accumulated
\cite{2NLS,MT,multihump,BSSDK}. Specifically, 
the soliton  \eqref{pqs} was proved to be stable \cite{MT,BSSDK}
and localised solutions with an arbitrary number of humps were determined
in addition to this fundamental soliton \cite{multihump}. 
By contrast,  the four-component  Schr\"odinger equation  \eqref{4NLS} 
 is not in the existing literature.

 An interesting property of equations \eqref{2NLS}  and \eqref{4NLS} is their conservativity.
In particular,  equation  \eqref{4NLS}  represents a Hamiltonian system with the Hamilton function
\begin{align*}
H    = \int \left[ |\dot p_1|^2 + |\dot p_2|^2 + |\dot q_1|^2 + |\dot q_2|^2    \phantom{|^{\frac12}}     \right.
\nonumber   \\       + |p_1|^4 + |p_2|^4                    + |q_1|^4 + |q_2|^4              \phantom{|^{\frac12}}             \nonumber \\  
-2 \left(  |p_1|^2 + |p_2|^2 + |q_1|^2 + |q_2|^2 \right)^2     \phantom{|^{\frac12}}  \nonumber \\   \left.  \phantom{|^{\frac12}}  
-4(p_1p_2q_1^* q_2^* + p_1^*p_2^*q_1 q_2)      \right] d T_1,   \phantom{|^{\frac12}}    
\end{align*}
where the overdot stands for 
$\partial / \partial T_1$.
 Equations \eqref{4NLS}  can be written as 
\begin{align*}
i \frac{\partial p_n}{\partial Z_2}= \frac{ \delta H}{\delta p_n^*},
\quad
i \frac{\partial q_n}{\partial Z_2}= \frac{ \delta H}{\delta q_n^*}
\quad (n=1,2),
\end{align*}
where 
$p_{1,2}^*$ are the momenta canonically conjugate to the coordinates $p_{1,2}$, and
$q_{1,2}^*$ are the momenta conjugate to $q_{1,2}$. 

Thus, despite the presence of gain and loss, 
the small-amplitude light pulses in the $\mathcal{PT}$-symmetric necklace obey Hamiltonian dynamics.

\section{Concluding remarks}
\label{Conclusions} 
\subsection{Conclusions}

When the coupled waveguides considered in this paper are linear and non-dispersive --- that is, when the system is modelled by 
the linear chain of $2N$ elements
  --- the complex modes are 
given by arbitrary linear combinations of eigenvectors of the $2N \times 2N$  matrix  \eqref{A5}. The addition of  the nonlinearity and dispersion
imposes nonlinear constraints on the coefficients of the admissible combinations.
We have classified  linear patterns that  persist in the nonlinear dispersive necklace.

One simple pattern arising in the necklace of $2N$ linear waveguides corresponds to $z$-independent illumination. 
The pattern consists of a linear combination of ${\vec v}^{(\alpha)}$ and ${\vec v}^{(N-\alpha)}$, two  eigenvectors pertaining to the repeated eigenvalue $\beta_\alpha$
(where $\alpha=1, ..., N-1$). 
A linear combination of ${\vec v}^{(\alpha)}$ and ${\vec w}^{(\alpha)}$  --- the eigenvectors associated with opposite eigenvalues ---
describes a periodic power oscillation between odd and even waveguides. (Here $\alpha$ may take any value from 1 to $N$.)
An odd-even blinking regime with the maximum waveguide power varying  along the necklace, is generated by  a combination of four eigenvectors:
${\vec v}^{(\alpha)}$, ${\vec w}^{(\alpha)}$, ${\vec v}^{(N-\alpha)}$ and ${\vec w}^{(N-\alpha)}$  ($\alpha=1, ..., N-1$).

The most interesting  types of structure result from combining  ${\vec v}^{(\alpha)}$ with ${\vec w}^{(N-\alpha)}$, or  ${\vec v}^{(\alpha)}$ with ${\vec v}^{(N)}$. 
With  either of these choices, 
 light propagates by switching from one guide to the next  in a corkscrew fashion.
A more complex, multiflash, spiral is associated with a pattern comprising four eigenvectors: 
${\vec v}^{(\alpha)}$, ${\vec w}^{(N-\alpha)}$, ${\vec v}^{(N)}$ and ${\vec w}^{(N)}$
($\alpha=1, ..., N-1$).

Our analysis of the nonlinear dispersive structures focussed on short pulses of light.
 Turning on the dispersion and nonlinearity, the configuration corresponding to 
the $z$-independent illumination transforms into a constellation of  $2N$ synchronised pulses.  The corresponding amplitudes of 
 supermodes are
 given by the soliton solutions of the one-  or two-component nonlinear Schr\"odinger equation
 (equation \eqref{A22} or \eqref{2NLS}, respectively).  On the other hand, the nonlinear dispersive counterpart of 
the odd-even oscillation consists of  a string of flashes. In that case,
the amplitudes of the eigenvectors constituting a two-supermode pattern satisfy the system \eqref{2NLS}
while in a four-supermode combination, 
 the amplitudes are solitons of the four-component  equation \eqref{4NLS}.

The  spiral patterns  in the  necklace of nondispersive linear waveguides persist as
 gyrating solitons of its nonlinear dispersive counterpart.
The gyrating soliton 
is a light pulse  that propagates along the fiber and circulates around the 
 necklace at the same time.
The soliton
 amplitudes of the spiral pattern  combining two eigenvectors --- 
 ${\vec v}^{(\alpha)}$ with ${\vec w}^{(N-\alpha)}$, or ${\vec v}^{(\alpha)}$ with ${\vec v}^{(N)}$ ---
 satisfy the system  \eqref{2NLS}.
The  helical structure
 involving {\it four\/}  supermodes
   gives rise to a multiflash gyrator: 
  a string of flashes with modulated brightness, revolving around the necklace as a whole.
  The amplitudes of the four eigenvectors ${\vec v}^{(\alpha)}$, ${\vec w}^{(N-\alpha)}$, ${\vec v}^{(N)}$, and ${\vec w}^{(N)}$.
  are given by the soliton solution of the 
  four-component nonlinear Schr\"odinger equation \eqref{4NLS}.

 Our numerical simulations indicate that some of the gyrating solitons are stable while 
 some other ones are weakly unstable. 
 
 The optical necklace  we considered in this paper was either 
  conservative (no gain no loss) or $\mathcal{PT}$-symmetric, 
where  lossy waveguides alternate with
waveguides with  gain.
Our perturbative construction of short-pulse solutions is equally applicable to both arrangements --- 
as long as the gain-loss coefficient in the nonhermitian necklace remains under the $\mathcal{PT}$-symmetry breaking threshold.

The nonhermitian necklace affords control opportunities unavailable in conservative arrays. We have shown that by varying the gain-loss coefficient 
one can change the length of the  pulse of light, its  velocity and sense of gyration.

 \subsection{Relation to earlier studies} 
 
It is appropriate to place our results in the context of existing literature on revolving light patterns.
 
  The authors of Ref \cite{Krolikowski} studied spatial solitons in the nonlinear hermitian necklace
 (equation \eqref{A1} without the $\partial_\tau^2 \psi_n$ term and with $\Gamma_n=0$). The localised structures of
Ref  \cite{Krolikowski} are travelling solitons of the one-dimensional discrete Schr\"odinger equation that were transplanted from 
 an infinite chain to a ring with a large but finite number  of sites. 
 Those structures are {\it not\/}  the gyrating solitons considered in this paper.
 The stationary light beams of Ref \cite{Krolikowski} are localised in $n$ whereas our gyrating solitons 
 are localised in the retarded time, $\tau$.

 Another class of circular patterns extensively covered in literature, comprises 
  azimuthons in the planar nonlinear Schr\"odinger equation \cite{azimuthons}.
 Azimuthons are ring-shaped  complexes of two-dimensional solitons
 revolving around a common centre. Unlike the gyrating solitons which are pulses travelling in waveguides, azimuthons are formed by
  stationary light beams   in  homogeneous media. Mathematically, the difference is that the azimuthon is a ring of several coexisting solitons
  involved in collective motion 
  whereas a gyrating soliton is a lone  pulse revolving around the necklace on its own.
  The azimuthon is not constrained by any lattice while the gyrating soliton requires a ring-shaped necklace to circulate.

Finally, we note parallels between the hermitian spiral patterns of the present study
and
 rotary beams in circular arrays reported in Ref \cite{Alexeyev}.
The principal difference between the system considered in Ref \cite{Alexeyev} and 
our equation \eqref{A1} with $\gamma=0$, is that the latter is nonlinear and takes into account  dispersion of pulses. 
These factors select particular spiral patterns that may form trajectories of the gyrating solitons. \\

\section*{Acknowledgments}
We thank Anton Desyatnikov, Boris Malomed and Sergei Turitsyn  for useful discussions.
This research was supported by the National Research Foundation of South Africa (grant 120844).


\end{document}